\documentclass{emulateapj}
\usepackage[dvips]{epsfig,color}
\usepackage{amsmath}

\newcommand{\eg}{{\it e.g.,}}
\newcommand{\ie}{{\it i.e.,}}
\newcommand{\etal}{{\it et al.}}

\newcommand{\ignore}[1]{\relax}
\newcommand{\simgt}{\hbox{\rlap{\raise 0.425ex\hbox{$>$}}\lower
0.65ex\hbox{$\sim$}}}
\newcommand{\simlt}{\hbox{\rlap{\raise 0.425ex\hbox{$<$}}\lower
0.65ex\hbox{$\sim$}}}

\DeclareMathAlphabet{\mathsfsl}{OT1}{cmss}{m}{sl}

\newcommand{\dif}{\mathrm{d}}

\newcommand{\vnab}{\ensuremath{\boldsymbol{\nabla}}}
\newcommand{\vvz}{\ensuremath{\boldsymbol{v_0}}}
\newcommand{\vvp}{\ensuremath{\boldsymbol{v_p}}}

\newcommand{\bcdot}{\ensuremath{\boldsymbol{\cdot}}}
\newcommand{\dfdt}{\ensuremath{\frac{\partial f}{\partial t}}}

\newcommand{\kb}{\ensuremath{k_{\rm B}}}
\newcommand{\vx}{\ensuremath{\boldsymbol{x}}}
\newcommand{\vv}{\ensuremath{\boldsymbol{v}}}
\newcommand{\va}{\ensuremath{\boldsymbol{a}}}

\shorttitle{}
\shortauthors{Barnes \& Williams}

\begin{document}

\title{Entropy Production in Collisionless Systems. II. Arbitrary
Phase-Space Occupation Numbers}
\author{Eric I. Barnes}
\affil{Department of Physics, University of Wisconsin --- La Crosse,
La Crosse, WI 54601}
\email{barnes.eric@uwlax.edu}
\author{Liliya L. R. Williams}
\affil{Minnesota Institue for Astrophysics, University of Minnesota,
Minneapolis, MN 55455}
\email{llrw@astro.umn.edu}

\begin{abstract}

We present an analysis of two thermodynamic techniques for determining
equilibria of self-gravitating systems. One is the Lynden-Bell entropy
maximization analysis that introduced violent relaxation. Since we do
not use the Stirling approximation which is invalid at small
occupation numbers, our systems have finite mass, unlike Lynden-Bell's
isothermal spheres. (Instead of Stirling, we utilize a very accurate
smooth approximation  for $\ln{x!}$.) The second analysis extends
entropy production extremization to self-gravitating systems, also
without the use of the Stirling approximation. In addition to the
Lynden-Bell (LB) statistical family characterized by the exclusion
principle in phase-space, and designed to treat collisionless systems,
we also apply the two approaches to the Maxwell-Boltzmann (MB)
families, which have no exclusion principle and hence represent
collisional systems. We implicitly assume that all of the phase-space
is equally accessible. We derive entropy production expressions for
both families, and give the extremum conditions for entropy
production. Surprisingly, our analysis indicates that extremizing
entropy production rate results in systems that have maximum entropy,
in both LB and MB statistics. In other words, both thermodynamic
approaches lead to the same equilibrium structures. 

\end{abstract}


\section{Introduction}\label{intro}

\subsection{Motivation}

Understanding how collisionless systems attain a specific mechanical
equilibrium state is fundamentally important to astrophysics.  For
example, the cold dark matter structures that exist around galaxies
are expected to fall into this class of system. Individual dark matter
constituents (whatever they may be) should evolve according to a
mean-field gravitational potential, free of the influence of
individual encounters.

The range of mechanical equilibria available to a collisionless system
is defined by the Jeans equation, which represents the condition that
no portion of the system experiences a net force. Unfortunately, the
Jeans equation admits an infinity of solutions.  Even if the mass
distribution is specified, there is an infinite set of acceptable
mechanical equilibria, each involving a different velocity
distribution.  For spherical systems, these velocity distributions
differ in their anisotropy profile that quantifies radial versus
tangential motion.  However, the question remains, how and/or why does
any one collisionless system evolve to its particular mechanical
equilibrium end-state, and what are the properties of such a state?

\subsection{Thermodynamic Approaches to the Problem}

The statistical mechanics description of thermodynamics provides one
path to obtaining the description of the final relaxed state.  A fully
relaxed system is the most statistically likely state of that system,
or the one with an entropy maximum.  In calculating the most likely
state it is implicitly assumed that all states of the system are
equally accessible.

Such an approach was taken by \citet{lb67} and applied to
self-gravitating collisionless systems, with the hope of explaining
the observed light distribution of elliptical galaxies. A
collisionless system can be thought of as a fluid in the 6D
phase-space of position and velocity. The `particles' in Lynden-Bell's
analysis are parcels of phase-space, \ie\ parcels of this fluid, and
so the distribution function (DF) representing the phase-space density
is defined in terms of energy per unit mass, not energy per particle.
The analysis resulted in a DF similar to the Fermi-Dirac case, but
with a different normalization.  \citet{lb67} argued that a
non-degenerate limit is appropriate for stellar systems, and thus
arrived at a DF similar to that of the Maxwell-Boltzmann case (an
exponential) which resulted in the isothermal sphere representing
thermal equilibrium. Since the isothermal sphere has an infinite
extent and mass, its emergence from the entropy maximization
procedure, which demanded a finite mass system, presented a
contradiction.  This apparent failure of entropy maximization was
puzzling, and it was often argued that such systems do not have states
of maximum entropy.  Some effort was made to investigate maximizing
entropy with additional constraints beyond mass, energy, and angular
momentum \citep{sb87,wn87}.  Other routes involving minimum energy
states of self-gravitating systems were also developed
\citep[\eg][]{a94}.

Recently, \citet{m96} \citep[based on earlier work by][]{s94} has
pointed out that the reason for the system's infinite mass was the use
of the Stirling approximation, $\ln{n!} \approx n\ln{n} -n$. In
contrast to systems usually treated in standard statistical mechanics,
self-gravitating systems can apparently have small phase-space
occupation numbers $n$, making Stirling a poor approximation.
Specifically, these systems have regions of phase-space or
energy-space that are nearly or completely unoccupied, such that $n$
will be small. Spatially, these regions can correspond to the center
of the potential as well as its outer edge. Using Maxwell-Boltzmann
statistics and the exact $\ln{n!}$, \citet{m96} has found a
distribution function from entropy maximization that is very similar
to \citet{k66} models.  \citet{hw10} have shown that the energy-space
occupation function $N(E)$ derived using a very accurate smooth
approximation to $\ln{n!}$ closely resembles the results of
collisionless $N$-body simulations \citep{whw10}.

\subsection{Brief Review of Statistical Representations}\label{stat}

From a statistical point of view, entropy is simply a measurement of
the number of states accessible to a particular system.  This
relationship is most commonly expressed quantitatively as,
\begin{equation}\label{s0}
S=\kb \ln{\Omega},
\end{equation}
where $\Omega$ is the number of accessible states and \kb\ is the
Boltzmann constant which serves to give entropy the correct
thermodynamic units.  As a result, counting procedures are key to
determining specific realizations of entropy.  \citet{lb67} discusses
how there are four counting types that lead to physically relevant
situations.  Bose-Einstein statistics follow from counting states for
indistinguishable particles that can co-habitate in the same state.  
When indistinguishable particles are not allowed to share states,
Fermi-Dirac (FD) statistics emerge.  Classical Maxwell-Boltzmann (MB)
statistics result from counting states available to distinguishable
particles that can share states.  Completing the symmetry, systems
where distinguishable `particles'---actually, parcels of
phase-space---cannot co-occupy states obey what has become known as
Lynden-Bell (LB) statistics (statistics `IV' in Lynden-Bell's original
notation).  Each type of statistics will produce different
representations of entropy, but we will focus on the two that deal
with classical, or distinguishable particles, namely, LB and MB.

We briefly recap the notation used in \citet{lb67} before proceeding
with our discussion. The six-dimensional position-velocity phase-space
(\vx,\vv) is the usual setting for determining the statistics.
Imagine phase-space to be divided into a very large number of nearly
infinitesimal parcels, called micro-cells, each having volume
$\varpi$.  Each micro-cell can either be occupied or unoccupied by one
of the $N$ phase-space elements of the system.  These elements can be
thought of as representing the fine-grained distribution function,
which has a constant density value $\eta$. Because collisionless
processes imply incompressibility of the fine-grained distribution
function, the phase elements cannot co-habitate.  If phase-space is
also partitioned on a coarser level so that some number $\nu$ of
micro-cells occupy a macro-cell, then we can discuss a coarse-grained
distribution function.  The volume of a macro-cell is then $\nu\varpi$
and the $i$th macro-cell contains $n_i$ phase elements.  We assume
that while the volume of a macro-cell is much larger than that of a
micro-cell, it is still very small compared to the full extent of
phase-space occupied by the system.  The number of ways of organizing
the $n_i$ elements into the $\nu$ micro-cells without co-habitation
is,
\begin{equation}\label{lbw0}
\frac{\nu!}{(\nu-n_i)!}.
\end{equation}
If the elements were allowed to multiply occupy micro-cells, as in MB
statistics, this number would be given by $\nu^{n_i}$.

To get the total number of accessible states, the possible ways to
distribute the $N$ phase elements into $n_i$ chunks must also be
included.  \citet{lb67} derives,
\begin{equation}\label{lbw}
\Omega_{\rm LB} = \frac{N!}{\prod_i n_i!} \times \prod_i 
\frac{\nu!}{(\nu-n_i)!}.
\end{equation}
In the MB case, the only change is that the factorial ratio in the
final product term is replaced by $\nu^{n_i}$. 

The LB case disallows two phase-space elements from inhabiting the
same phase-space location, so it explicitly takes into account the
incompressibility of a collisionless fluid. Since we are primarily
interested in dark matter halos, LB is the natural case to consider.
For completeness, and for the sake of having a comparison, we also
treat the MB case, in the Appendix.

We argue that the lack of an exclusion principle in the MB case is
equivalent to allowing collisions between particles.  In a collisional
system, particles from distant phase-space locations can be scattered
into any other phase-space location, thereby increasing the
phase-space density at the latter location. In principle, there is no
limit to how high the density can get through such scatterings.  In
practice, the phase-space density probably can not become very high at
most locations, but it can be higher than the original fine-grained
DF.  We note that it is common to use MB to represent collisional
systems.  For example, \citet{m96} argues that it is the correct
statistics to use for globular clusters where the relatively small
number of stars allows the cluster to relax through two-body
interactions.  It then makes sense that the energy distribution that
the cluster will arrive at will be the same as that in a cloud of gas,
which relaxes through collisions between molecules.  In the
non-degenerate limit, when the micro-cells are very sparsely populated
and the density of the coarse-grained distribution function is very
dilute compared to that of the fine-grained function, both LB and MB
distribution functions, and hence density profiles, will look the
same. 

\subsection{This Work}\label{twork}

In this paper we explore two possible approaches to deriving the final
equilibrium state of self-gravitating systems, for each of the two
types of statistics, LB and MB. In both, we use a very accurate,
smooth approximation for $\ln{x!}$ valid for arbitrary occupation
numbers $x$.  However, the price we pay for this improved
approximation is the loss of analytic solutions.

The first approach assumes that the final state is the maximum entropy
state, an assumption that was first used in the context of
self-gravitating systems in 1950's \citep{o57}. The second approach,
again in the context of self-gravitating systems, was first taken by
\citet{bw11}; it posits that the final state corresponds to the
extremum of entropy production rate. 

We explore extremizing entropy production because it has not been
established beyond a doubt that real or computer simulated systems do
fully relax to maximum entropy states. It is possible that their
steady-state configurations do not correspond to maximum entropy.
Prior work on thermal non-equilibrium systems, but not in
astrophysical contexts, suggests that stationary states---like
mechanical equilibrium---occur when entropy production is extremized
\citep[\eg][]{p61,j80,dgm84,g08}. We investigate their applicability
to self-gravitating systems in mechanical equilibrium.

The new aspect in the present paper is that we use a very accurate
approximation for $\ln{x!}$, unlike our previous paper that assumed
the Stirling approximation. As \cite{m96} has shown, replacing
Stirling with an accurate approximation (i) results in systems with
finite total mass and energy, and (ii) significantly changes the
structure of the systems. 

In all, we present four derivations; entropy maximization for the LB
and MB statistics are covered in Sections~\ref{lbsmax} and
~\ref{smax}.  Extremization of entropy production for LB and MB
statistics are carried out in Sections~\ref{lbsigma} and
~\ref{sigmax}, respectively.  We develop expressions for the
relaxation functions (see \S~\ref{recap}), and use these to better
understand the evolution of coarse-grained distribution function.  We
compare our results with analogous versions of entropy production
derived using the standard Stirling approximation in \citet{bw11}.

\begin{figure}
\plotone{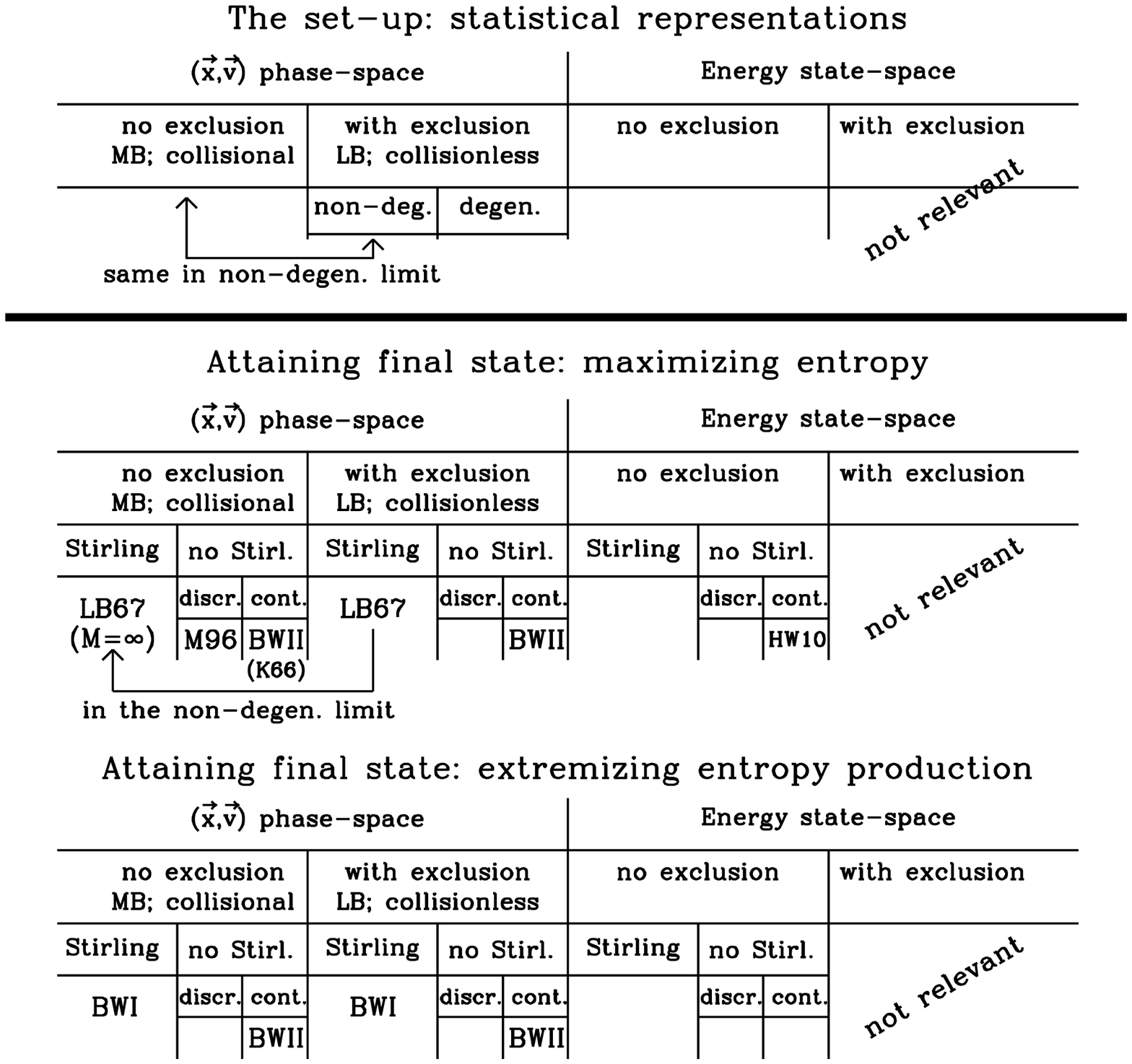}
\figcaption{A schematic summary of the various statistical mechanical 
approaches to self-gravitating systems. See Section~\ref{twork} for 
explanation.
\label{tblfig}}
\end{figure}

Figure~\ref{tblfig} puts the present paper (BWII in the figure) in
context.  It is a schematic summary of the various statistical
mechanical approaches to self-gravitating systems. The possible ways
to frame the problem appears at the top of the figure; one can
formulate the problem in either the regular phase-space, or the energy
space. Below the thick horizontal line we show the two different
routes for attaining the final steady-state state: maximizing entropy,
and extremizing entropy production. Once these choices are made one
has to decide whether small occupation number regime will be important
or not, and hence whether to use the Stirling approximation for $\ln
x!$, or not.  In the latter case, one must then decide whether to use
the discrete (``discr.'') step-like, i.e. exact version of $\ln x!$,
or to approximate it with some smooth function (``cont.'') which
remains very accurate down to small $x$. Note that HW10 and the
present paper use different but similar approximations. There is no
physical reason to introduce the exclusion principle in the energy
state-space, hence the corresponding regions are marked as ``not
relevant''. The bottom entries of some columns in the table contain
names of papers where the corresponding options were considered. K66
in parentheses below BWII means that \citet{k66} results are nearly
identical to ours.  $M=\infty$ under LB67 means that \citet{lb67}
final result, the isothermal sphere, had infinite mass.

Much of the background material for this work may be found in
\citet{bw11}, and we briefly summarize these previously obtained
results in Section~\ref{recap}.  

\section{Summary of Results for Large Occupation Numbers}\label{recap}

\citet{bw11} have investigated entropy production in self-gravitating
systems described by MB and LB and have developed expressions for the
entropy production, $\sigma$ for both the MB,
\begin{equation}\label{sigmb0}
\sigma_{\rm MB}=
-\frac{\kb}{\varpi \eta} \int \Gamma \left[ 
\ln{\left(\frac{f}{\eta}\right)} +1 -\ln{N} \right] \: \dif\vv,
\end{equation}
and LB cases,
\begin{equation}\label{siglb0}
\sigma_{\rm LB}=
-\frac{\kb}{\varpi \eta} \int \Gamma
\left[\ln{\left(\frac{f}{\eta-f}\right)}-C \right] \: \dif\vv,
\end{equation}
where constant $C$ in Equation~\ref{siglb0} is 
$(\ln{N}-1)+(1/N)\sum_i \nu\ln{\nu}$.
In these expressions, $f$ is the coarse-grained distribution function
and $\Gamma$ is the relaxation function and forms the right-hand side
of the Boltzmann equation,
\begin{equation}\label{boltzeqn}
\dfdt + \boldsymbol{v} \bcdot \vnab f +
\va \bcdot \vnab_v f = \Gamma(f).
\end{equation}
If one were to assume, for a moment, that in the above equation $f$ is
a fine-grained DF, then the right hand side would be zero for
collisionless systems. In other words $\Gamma=0$, which means that on
fine-grained scales there is no change in, or production of entropy;
in collisionless systems, entropy is fixed throughout evolution.  Let
us be clear, we are not advocating for a specific process as a source
for $\Gamma$, like collisions in a gas.  The relaxation function
simply describes the Lagrangian time rate of change of the
distribution function.  Returning to our case where $f$ represents the
coarse-grained DF, Equation~\ref{boltzeqn} states that entropy is
produced [and as we argue in \cite{bw11} it happens even in systems
that have attained macroscopic steady-state] because on microscopic
scales the fine-grained DF continues to wind and twist, which when
combined with coarse-graining, gives rise to non-zero entropy change.
From a different starting point, \citet{c98} develops an expression
for the right-hand side of Equation~\ref{boltzeqn} in terms of a
``diffusion current'' that relates to correlations between
fluctuations in the fine-grained distribution function.  While that
work details the makeup of this diffusion current, we simply focus
on the broad behavior of the relaxation function.

We find extremum entropy production conditions by setting the
variation of entropy production, $\delta \sigma$ equal to zero.  This
operation gives expressions for the relaxation function.  For the MB
case,
\begin{equation}\label{gmb0}
\Gamma_{\rm MB}(f)=\frac{(1-\ln{N})\Gamma_{\rm MB}(f=\eta)}
{\ln{(f/\eta)} +1 -\ln{N}}.
\end{equation}
The LB relaxation function is slightly more complex,
\begin{equation}\label{glb0}
\Gamma_{\rm LB}(f)=\frac{-C\Gamma_{\rm LB}(f=\eta/2)}
{\ln{[f/(\eta-f)]}-C}.
\end{equation}
Like \citet{lb67}, the \citet{bw11} work assumes that the large $n$
Stirling approximation is valid for the systems being investigated.
Here, we will be deriving relations analogous to
Equations~\ref{sigmb0}, \ref{siglb0}, \ref{gmb0}, and \ref{glb0}, but
using a very accurate approximation, after discussing the results of
entropy maximization below.

\section{Results for Arbitrary Occupation Numbers}

\begin{figure}
\plotone{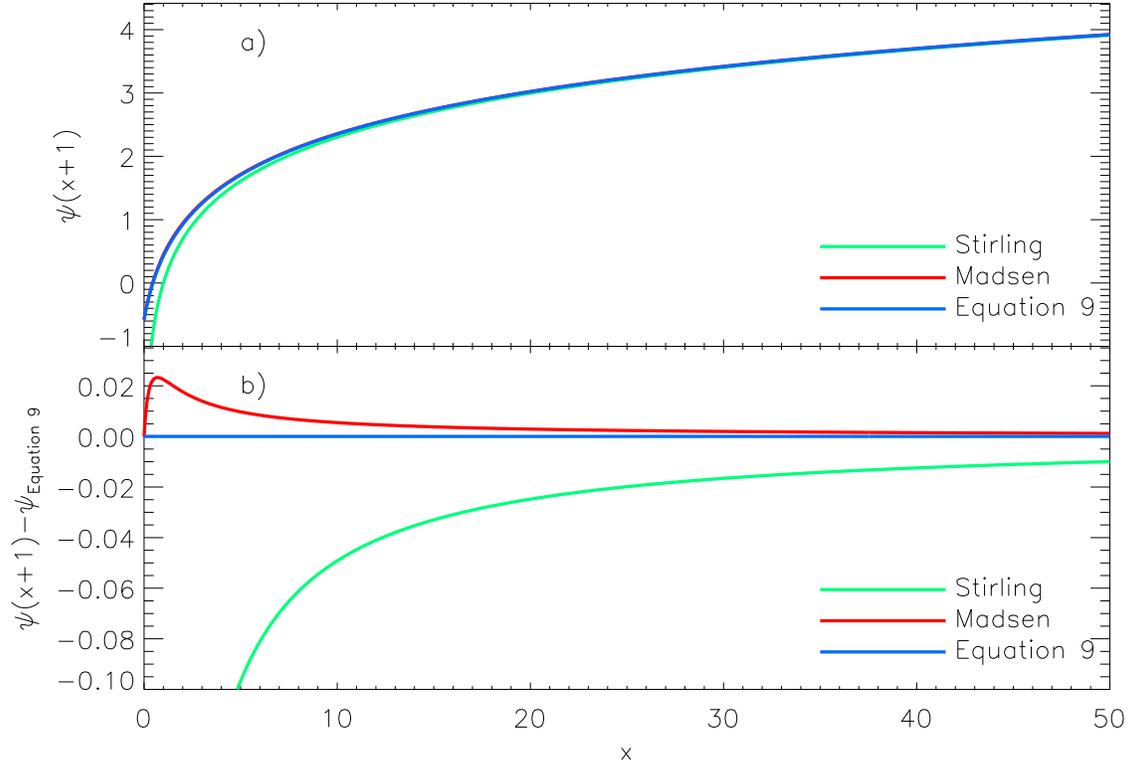}
\figcaption{Comparing the Stirling, \cite{hw10} and
Equation~\ref{approx} approximations to $\ln{x!}$ as functions of $x$.
The function $\psi(x+1) \equiv \dif \ln{x!}/\dif x$ is plotted.  Note
the divergent behavior of the Stirling approximation as $x \rightarrow
0$.  (a) The raw $\psi$ functions for the three approximations are
shown.  The \cite{hw10} line is very nearly covered by the
Equation~\ref{approx} line.  (b) Differences between the various
approximations and Equation~\ref{approx}.  The differences between the
\cite{hw10} and Equation~\ref{approx} approximations are much smaller
than those between either approximation and the Stirling
approximation.  Also, the \cite{hw10} and Equation~\ref{approx} values
coincide for $x=0$.
\label{appfig}}
\end{figure}

The approximation that we utilize in this work is,
\begin{equation}\label{approx}
\ln{x!}=(x+\frac{1}{2})\ln{(x+1)} - x + \frac{\ln{2\pi}}{2} +
\lambda_{0,x}
\end{equation}
where 
\begin{equation}
\lambda_{0,x}=-\frac{(x^2 + 2x + \frac{287}{288})}{(x^2 +
\frac{25}{12}x + \frac{13}{12})}.
\end{equation}
In the large $x$ limit, this reduces to the usual Stirling approximation 
$\ln{x!}=x\ln{x}-x$.  Our approximation is slightly different from the one 
in \citet{hw10}. Comparisons between the Stirling approximation, \citet{hw10}
approximation, and Equation~\ref{approx} are shown in Figure~\ref{appfig}.  
The plots illustrate the function $\psi(x+1) \equiv \dif \ln{x!}/\dif x$ for 
the three cases.  The HW10 and Equation~\ref{approx} approximations are 
nearly identical. Most importantly, these two approximations are well-behaved 
at $x=0$, unlike the Stirling approximation.


\subsection{Entropy Maximization}\label{lbsmax}

In this section, we follow the overall path taken in \citet{lb67} to
determine the description of entropy for the LB statistics family.
Briefly, counting arguments for phase-space macro-cell occupation are
combined with the statistical definition of entropy to give specific
representations.  Variations in entropy assuming constant mass and
energy are then set to zero in order to determine the necessary
distribution functions.

If one demands that multiple phase-space elements cannot
simultaneously occupy micro-cells, the multiplicity of states
(Equation~\ref{lbw}) combined with the definition of entropy results
in,
\begin{equation}
S_{\rm LB} = \kb \left[ \ln{N!} - \sum_i \ln{n_i!} + \sum_i \ln{\nu !}
- \sum_i \ln{(\nu - n_i)!} \right],
\end{equation}
where again the summations run over the number of macro-cells.  We
will assume that both $N$ and $\nu$ are much larger than 1, so that
Stirling's approximation is valid for use in the first and third
terms.  However, for the $n_i$ and $\nu - n_i$ terms we will use our
improved approximation,
\begin{equation}\label{approxn}
\ln{n_i!}=(n_i+1/2)\ln{(n_i+1)} - n_i + \frac{\ln{2\pi}}{2} +
\lambda_{0,n_i}.
\end{equation}
Note that this usage does not require either $n_i$ or $\nu - n_i$ to
actually be a small value, rather this keeps the accounting accurate
in the event that they do become small.  If these values are always
large, there will be no difference from the Stirling approximation.
The entropy expression now reads,
\begin{eqnarray}
S_{\rm LB} & = & S_{\rm LB,0} - \kb \sum_i \left[
(n_i+1/2)\ln{(n_i+1)} + \right. \nonumber \\
& & (\nu -n_i +1/2)\ln{(\nu-n_i+1)} + \lambda_{0,n_i} +
\nonumber \\
& & \left. \lambda_{0,(\nu-n_i)} \right],
\end{eqnarray}
where $S_{\rm LB,0} = \kb [ N\ln{N} - N + M(\nu \ln{\nu} -
\ln{2\pi})]$, and $M$ is the total number of macro-cells.

Transforming from the discrete macro-cell occupation number $n_i$ to
the continuous coarse-grained distribution function $f$, the entropy
becomes,
\begin{eqnarray}\label{slb0}
S_{\rm LB} & = & S_{\rm LB,0} - \frac{\kb}{\nu \varpi} \iint \left[
\left( \frac{\nu f}{\eta} + \frac{1}{2} \right) \ln{\left(\frac{\nu
f}{\eta} + 1 \right)} + \right. \nonumber \\
 & & \left( \nu - \frac{\nu f}{\eta} +
\frac{1}{2}\right) \ln{\left( \nu -\frac{\nu f}{\eta} +1 \right)} +
\lambda_{0,\nu f/\eta} + \nonumber \\
& & \left. \lambda_{0,\nu-\nu f/\eta} \right] \: \dif\vx \: \dif\vv.
\end{eqnarray}
Taking the variation of this entropy expression to be zero, with
constant mass and energy constraints, leads to the following
condition,
\begin{eqnarray}\label{slb1}
\lefteqn{\ln{(F+1)} + \frac{(F+1/2)}{(F+1)} - \ln{(\nu -F+1)} -}
\nonumber \\ & & \frac{(\nu -
F+1/2)}{(\nu - F + 1)} + \nonumber \\
 & & \frac{\dif \lambda_{0,F}}{\dif F} +
\frac{\dif \lambda_{0,\nu-F}}{\dif F} +\mu + \beta \epsilon = 0,
\end{eqnarray}
where $F=\nu f/\eta$ is a scaled coarse-grained distribution function
and $\mu$ and $\beta$ are undetermined multipliers associated with
mass and energy conservation, respectively.  The $\epsilon$ term is
the specific energy of a phase element located at position $\vx$ with
velocity $\vv$, $\epsilon=v^2/2 + \Phi$.  The derivative of the
$\lambda$ function is,
\begin{equation}\label{lamderiv}
\frac{\dif \lambda_{0,F}}{\dif F} = \frac{-(F+1)}{(F+600/576)}.
\end{equation}
After substituting for these $\lambda$ derivatives 
and combining terms we have,
\begin{eqnarray}\label{slb2}
\lefteqn{\ln{\left[ \frac{F+1}{\nu -F+1} \right]} + \frac{(F-\nu/2)}{(F+1)(\nu
-F+1)} -} \nonumber \\
 & & \frac{2F^2 - 2F\nu -(1+600/576)\nu - 600/288}{F^2 -F\nu -
(600/576)\nu +(600/576)^2} + \nonumber \\
 & & \mu + \beta \epsilon = 0.
\end{eqnarray}
The inelegant ratio terms in this expression are both symmetric about
$F=\nu/2$.  The second term on the left-hand side of
Equation~\ref{slb2} has values of $-1/2$ when $F=0$, $0$ when
$F=\nu/2$, and $1/2$ when $F=\nu$ (the maximum value of $F$ for the LB
case).  The third term on the left-hand side of Equation~\ref{slb2} is
a nearly constant function with a value very close to $-2$ for $0 \le
F \le \nu$.

\begin{figure}
\plotone{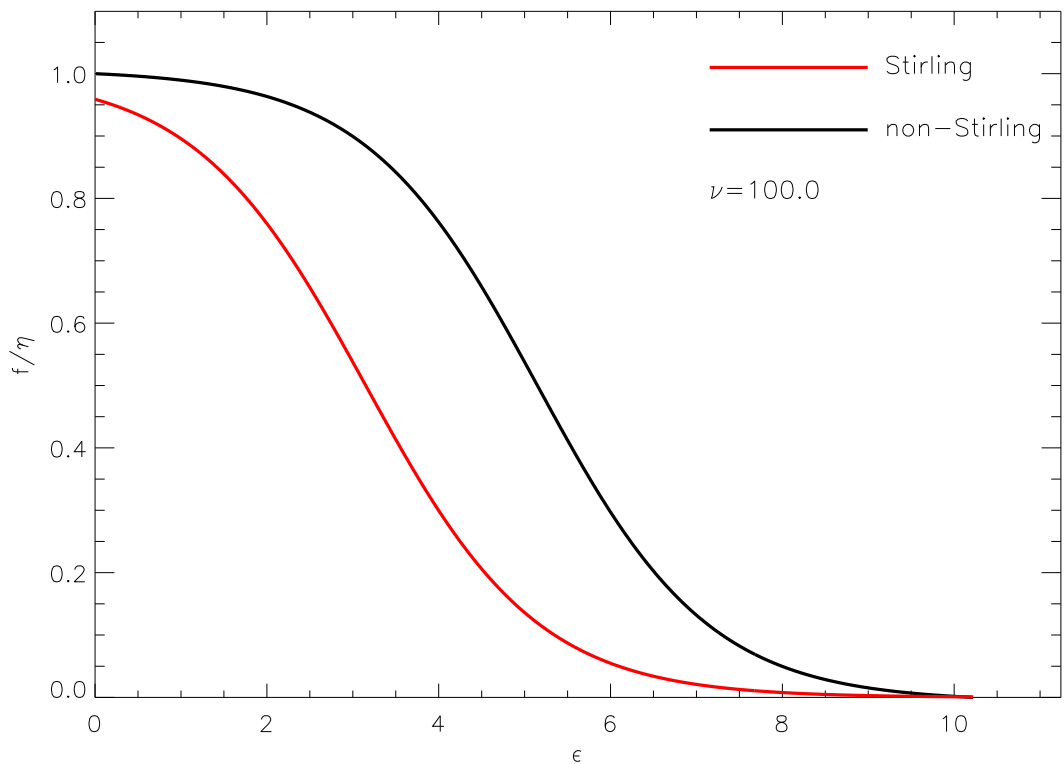}
\figcaption{The Lynden-Bell distribution function resulting from
the non-Stirling approximation.  The Lynden-Bell function derived
using the Stirling approximation is also shown.
\label{f_lb}}
\end{figure}

We have not attempted to find an analytic solution for $F$.  Graphical
solutions of Equation~\ref{slb2} for a series of $\epsilon$ values
produces the picture of $f/\eta$ seen in Figure~\ref{f_lb}. The
overall character of the distribution function is the same in the
non-Stirling and Stirling versions, and both functions are very
similar to Fermi-Dirac distribution,
\begin{equation}
f_{\rm FD}=\frac{\exp{-(\mu+\beta \epsilon)}}{1+\exp{-(\mu+\beta
\epsilon)}}.
\end{equation}

This distribution function of Equation~\ref{slb2} can be transformed
into a density distribution using the Poisson equation and the fact
that
\begin{equation}
\rho(r)=\int f(r,v) \dif\vv = 4\pi \int f(\epsilon)
\sqrt{2[\epsilon-\Phi(r)]} \dif \epsilon.
\end{equation}
Note that this procedure imposes an isotropic velocity distribution
for the system.

\begin{figure}
\plotone{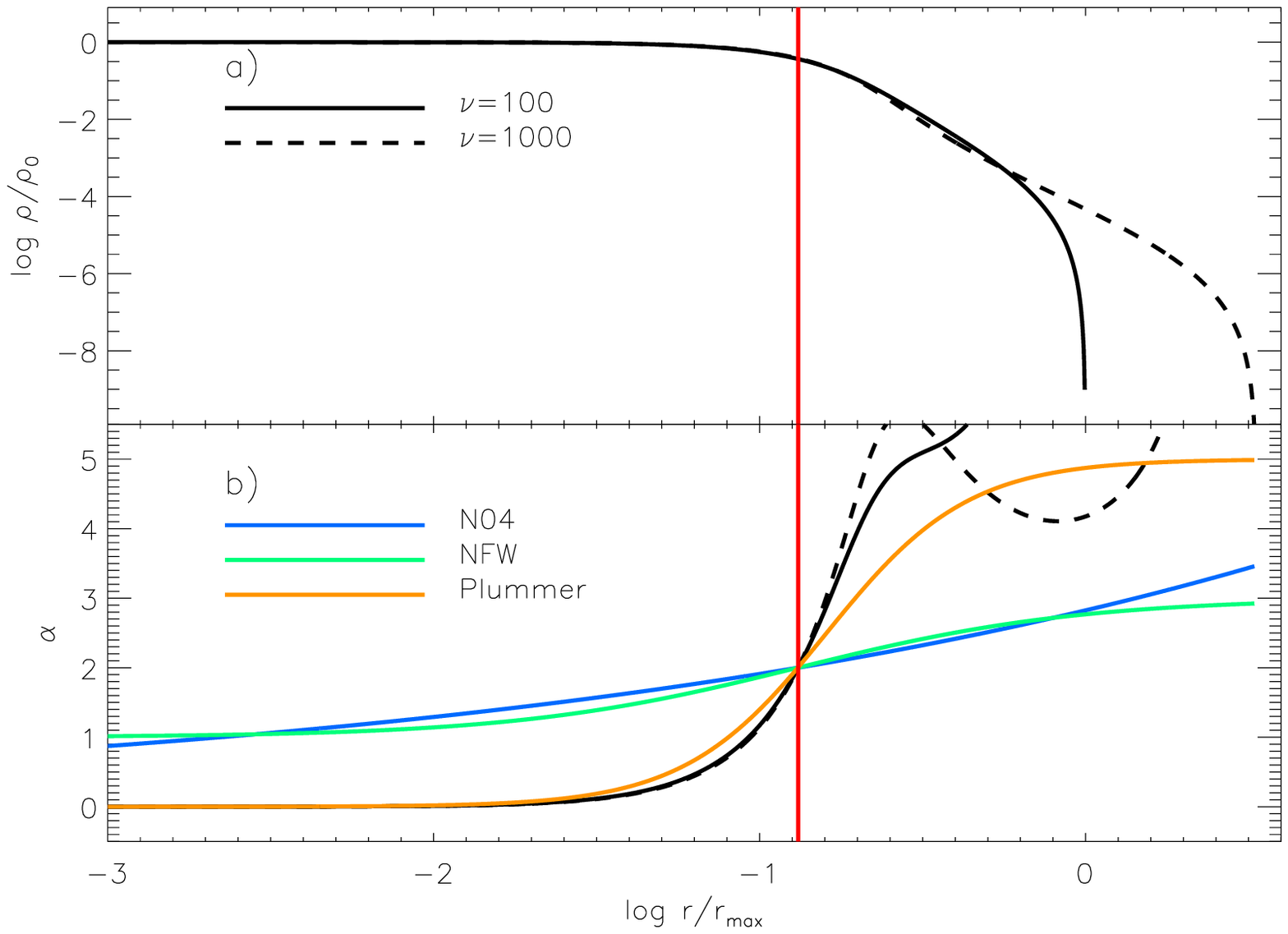}
\figcaption{(a) Logarithmic density and (b) the corresponding
$\alpha=-\dif \log{\rho}/\dif \log{r}$ versus logarithmic scaled
radius for the Lynden-Bell case.  The thick solid black lines
correspond to models with $\nu=100$ while the thick dashed black lines
represent models with $\nu=1000$.  The vertical lines mark where the
logarithmic slope of the $\nu=100$ model is isothermal ($\alpha=2$).
The $\alpha$ profiles for NFW, \cite{n04}, and Plummer models are also
shown.  The profiles for the $\nu=1000$ model have been horizontally
shifted so that the location where $\alpha=2$ coincides with the other
models.  The central cusp of the NFW and \cite{n04} models ($\alpha
\rightarrow 1$ as $r \rightarrow 0$) differs markedly from the core
present in the Lynden-Bell case ($\alpha \rightarrow 0$ as $r
\rightarrow 0$).  The presence of the core in the Lynden-Bell case is
independent of $\nu$.
\label{rho_lb}}
\end{figure}

\begin{figure}
\plotone{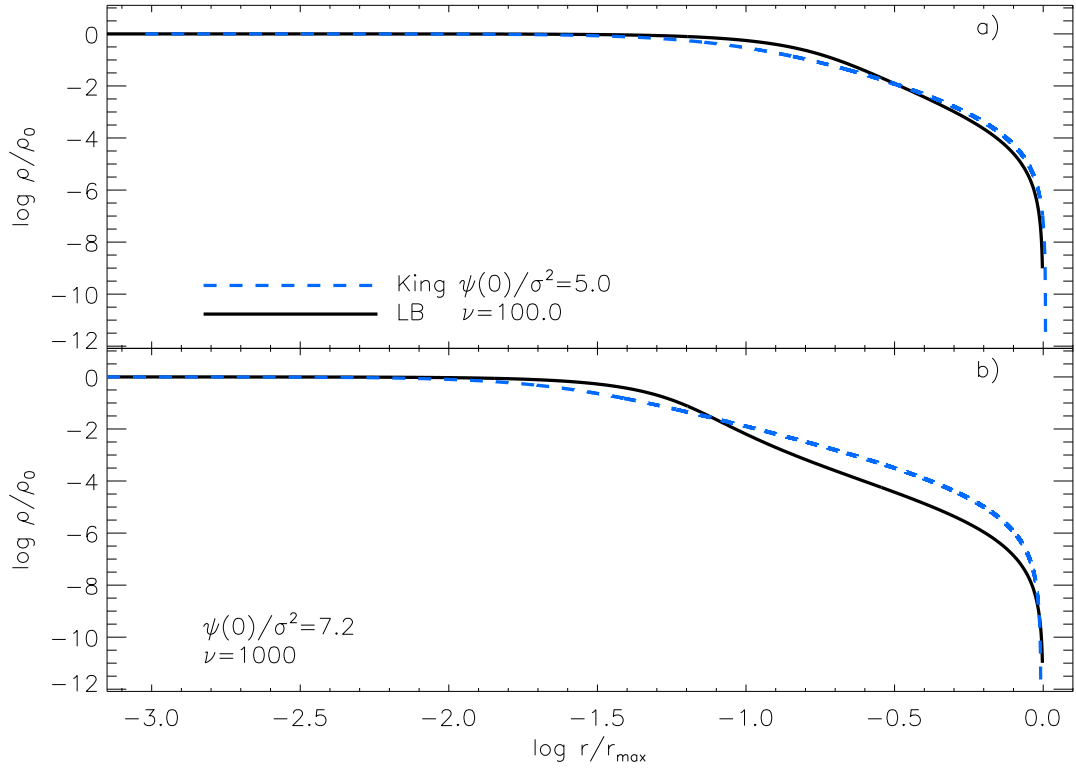}
\figcaption{Comparisons between Lynden-Bell (solid) and King (dashed)
density profiles for different $\nu$ values; (a) $\nu=100$ and (b)
$\nu=1000$.  The King scale factors correspond to those used in
Figure~\ref{kingcomp} for the same $\nu$ values.  The shapes of the
curves cannot be brought into agreement for any scale factor values.
\label{lbcomp}}
\end{figure}

The density and logarithmic density slope corresponding to the
non-Stirling function are presented in Figures~\ref{rho_lb} and
\ref{lbcomp}, and will be discussed further in Section~\ref{conc1a}.

\subsection{Entropy Production Extremization}\label{lbsigma}

Just because it is possible to describe thermal equilibrium states for
collisionless self-gravitating systems, it does not follow that real
systems (either physical or simulated) must achieve them in a Hubble
time.  It is possible that real systems incompletely relax, leaving
them in a thermal non-equilibrium, but long-lived stationary state.
\citet{m96} also points out that in very slowly evolving systems, the
maximization of entropy is only temporary.  This leads us to infer
that the production of entropy may be a useful quantity for discussing
quasi-equilibria of collisionless systems.  With that in mind, we now
proceed to develop the conditions required for a thermal
non-equilibrium state to be stationary.  Specifically, we find an
expression for entropy production in the LB statistical family and
then extremize it.  (As for the preceding section, an analogous
derivation for the Maxwell-Boltzmann statistics may be found in
Appendix~\ref{sigmax}.)  As discussed in Section~\ref{twork}, existing
work on thermal non-equilibrium in non-astrophysical settings suggests
that stationary states occur when entropy production is either maximum
or minimum \citep[][and references therein]{dgm84,g08}.

From Equation~\ref{slb0} the entropy density in the Lynden-Bell case 
can be written as,
\begin{eqnarray}\label{lbsdens}
\lefteqn{\rho s_{\rm LB} = -\frac{\kb}{\nu \varpi} \int \Bigg[
(F+1/2)\ln{(F+1)} -} \nonumber \\ & & (\nu-F+1/2)\ln{(\nu-F+1)} +
\nonumber \\
 & & \lambda_{0,F} +
\lambda_{0,\nu-F} - F\frac{S_{\rm LB,0}}{N\kb} \Bigg] \: \dif\vv,
\end{eqnarray}
where, as before, $F \equiv \nu f/\eta$.  

Taking a partial time derivative of Equation~\ref{lbsdens} results in
\begin{eqnarray}\label{lbk1}
\lefteqn{\frac{\partial}{\partial t}(\rho s_{\rm LB}) = -\frac{\kb}{\nu
\varpi} \int \frac{\partial F}{\partial t} \left[ \ln{(F+1)} +
\right.} \nonumber \\
& & \frac{F+1/2}{F+1} - \ln{(\nu-F+1)} +
\frac{\nu-F+1/2}{\nu-F+1} + \nonumber \\
& & \left. \frac{\partial \lambda_{0,F}} {\partial F}
+ \frac{\partial \lambda_{0,\nu-F}}{\partial F} - C \right] \:
\dif\vv,
\end{eqnarray}
where $C=S_{\rm LB,0}/N\kb$ is a constant.

Upon substituting $\partial F/\partial t$ from the Boltzmann equation 
into Equation~\ref{lbk1}, we get a lengthy expression,
\begin{eqnarray}\label{lbk2}
\lefteqn{
\frac{\partial}{\partial t}(\rho s_{\rm LB}) = } \nonumber \\
& & -\frac{\kb}{\nu \varpi} \int (-\vv \bcdot \vnab F) \Bigg[
\ln{(F +1)} + \frac{(F+1/2)}{(F+1)} - \nonumber \\
& & \ln{(\nu-F+1)} + \frac{\nu-F+1/2}{\nu-F+1} +
\frac{\partial \lambda_{0,F}}{\partial F} + \frac{\partial
\lambda_{0,\nu-F}}{\partial F} - C \Bigg] \: \dif\vv -
\nonumber \\
 & & \frac{\kb}{\nu \varpi} \int (-\va \bcdot \vnab_v F) \Bigg[
\ln{(F +1)} + \frac{(F+1/2)}{(F+1)} - \nonumber \\
& & \ln{(\nu-F+1)} + \frac{\nu-F+1/2}{\nu-F+1} + 
\frac{\partial \lambda_{0,F}}{\partial F} + \frac{\partial
\lambda_{0,\nu-F}}{\partial F} - C \Bigg] \: \dif\vv -
\nonumber \\ 
& & \frac{\kb}{\nu \varpi} \int \gamma \Bigg[ 
\ln{(F +1)} + \frac{(F+1/2)}{(F+1)} - \nonumber \\
& & \ln{(\nu-F+1)} + \frac{\nu-F+1/2}{\nu-F+1} + 
\frac{\partial \lambda_{0,F}}{\partial F} + \nonumber \\
& & \frac{\partial \lambda_{0,\nu-F}}{\partial F} - C \Bigg] \: \dif\vv,
\end{eqnarray}
where $\gamma=\nu \Gamma/\eta$.

The first term on the
right-hand side of Equation~\ref{lbk2} can be transformed into $-\vnab
\bcdot \rho s_{\rm LB} \vv$.  The acceleration-dependent terms in
Equation~\ref{lbk2} all disappear under the assumptions that the
distribution function is even in $v$ and disappears when $v=v_{\rm
max}$.  The final representation of the time rate of change of the
entropy density is,
\begin{eqnarray}\label{lbkk3}
\lefteqn{
\frac{\partial}{\partial t}(\rho s_{\rm LB}) = } \nonumber \\
& & -\frac{\kb}{\nu \varpi} \vnab \bcdot \int [ (F+1/2)
\ln{(F+1)} - \nonumber \\
& & (\nu-F+1/2)\ln{(\nu-F+1)} + \lambda_{0,F} +
\lambda_{0,\nu-F} -FC ] \vv \: \dif\vv - \nonumber
\\
& & \frac{\kb}{\nu \varpi} \int \gamma(F)
\Bigg[\ln{(F+1)} + \frac{F+1/2}{F+1} - \ln{(\nu-F+1)} + \nonumber \\
& & \frac{\nu-F+1/2}{\nu-F+1} + \frac{\dif \lambda_{0,F}}{\dif F} +
\frac{\dif \lambda_{0,\nu-F}}{\dif F} -C \Bigg] \: \dif\vv.
\end{eqnarray}
Again assuming the velocity field to be composed of mean and
peculiar components $\vv = \vvz + \vvp$, we draw a correspondence
between the terms in,
\begin{eqnarray}\label{lbkk4}
\lefteqn{
\frac{\partial}{\partial t}(\rho s_{\rm LB}) = 
-\vnab \bcdot \rho s_{\rm LB} \vvz - } \nonumber \\
& & \frac{\kb}{\nu \varpi} \vnab \bcdot \int [ (F+1/2)
\ln{(F+1)} - \nonumber \\ & & (\nu-F+1/2)\ln{(\nu-F+1)} + 
\lambda_{0,F} +
\lambda_{0,\nu-F} -FC ] \vvp \: \dif\vv - \nonumber \\
& & \frac{\kb}{\nu \varpi} \int \gamma 
\Bigg[\ln{(F+1)} + \frac{F+1/2}{F+1} - \ln{(\nu-F+1)} + \nonumber \\
& & \frac{\nu-F+1/2}{\nu-F+1} +
\frac{\dif \lambda_{0,F}}{\dif F} +
\frac{\dif \lambda_{0,\nu-F}}{\dif F} -C \Bigg] \: \dif\vv.
\end{eqnarray}
and those in the continuous version of the entropy density 
evolution equation, \citep[Equation 6 in][]{bw11}, 
\begin{equation}\label{s6}
\frac{\partial}{\partial t}(\rho s) = -\vnab \bcdot
(\boldsymbol{\Sigma} + \rho s \vvz) + \sigma.
\end{equation}

The entropy flux due to random motions $\boldsymbol{\Sigma}$ is given
by the integral in the second term on the right-hand side of
Equation~\ref{lbkk4}.  The remaining term then makes up the entropy
production for the system,
\begin{eqnarray}\label{lbsprod}
\sigma_{\rm LB} & = &
-\frac{\kb}{\nu \varpi} \int \gamma
\Bigg[ \ln{(F+1)} + \frac{F+1/2}{F+1} - \ln{(\nu-F+1)} + \nonumber \\
& & \frac{\nu-F+1/2}{\nu-F+1} + \frac{\dif \lambda_{0,F}}{\dif F} +
\frac{\dif \lambda_{0,\nu-F}}{\dif F} -C \Bigg] \: \dif\vv.
\end{eqnarray}
As expected, the relaxation function determines the entropy production
rate for the system.

We find that the condition for the entropy production term in
Equation~\ref{lbsprod} to be extremized is,
\begin{eqnarray}
\lefteqn{\Gamma_{\rm LB}(f)=Q/\left[ \ln{\left( \frac{F+1}{\nu-F+1}
\right) } + \frac{F+1/2}{F+1} + \right.} \nonumber \\
& & \left. \frac{\nu-F+1/2}{\nu-F+1} + \frac{\dif
\lambda_{0,F}}{\dif F} + \frac{\dif \lambda_{0,\nu-F}}{\dif F} - C
\right],
\end{eqnarray}
where the integration constant is defined by,
\begin{equation} 
Q=\gamma_{\rm LB}(F=\nu/2) \left[ \frac{\nu+1}{\nu/2+1}
- \frac{2(\nu+2)}{\nu+600/288} - C \right],
\end{equation}
and the $\lambda$ derivatives are the same as those used in going from
Equation~\ref{slb1} to \ref{slb2}.  This relaxation function is shown
in Figure~\ref{lbrelax} and is very similar to the corresponding
function from \citet{bw11}, their Figure~2.

\section{Summary and Discussion}\label{conclude}

In an attempt to better understand the evolution of self-gravitating
collisionless systems, we have re-investigated a standard statistical
mechanics approach to finding equilibria (entropy maximization), and
continued to develop a new approach (extremization of entropy
production), first applied to self-gravitating systems by
\citet{bw11}.  Entropy production in non-equilibrium steady-state
systems has been previously investigated and found to be useful in
several non-astrophysical systems, and so it is interesting to ask if
the principle is relevant in gravitational systems.

Both of our approaches use a very accurate  approximation for $\ln
n!$, and not the standard Stirling formula valid only for large
occupation numbers. It has been recently shown that  using an
approximation that reflects correct behavior for small $n$, and the
principle of entropy maximization leads to density distributions that
resemble globular clusters \citep[King profiles;][]{m96}, and
simulated pure dark matter halos \citep{wh10,whw10}. Both types of
systems have finite mass and energy, in compliance with the entropy
maximization constraints.

\subsection{Results of Entropy Maximization}\label{conc1a}

Entropy maximization using the LB and MB statistics produces the
density profiles and logarithmic profiles slopes, $\alpha= - \dif
\log{\rho}/\dif \log{r}$ depicted in Figures~\ref{rho_lb} and
\ref{rho_mb}, respectively. 

We argue that the LB case accurately represents collisionless systems
because co-habitation of the phase-space elements is prohibited.
However, the density profiles for any value of $\nu$ is very different from 
the two cosmological models, Navarro-Frenk-White (NFW) \citep{nfw96} and
N04 \citep{n04}, which are fits to the results of cosmological
$N$-body simulations.  In contrast to the latter, the LB density
profiles have a flat central density slope, while at larger radii
$\alpha$ has a steeper rise than even that of the Plummer profile
\citep{s83,p11,bt87} which is an example of a polytropic system.  In
Figure~\ref{lbcomp}, we compare the profiles to those derived in
Section~\ref{smax}, which are nearly identical to King models; LB
profiles are poor matches to King models. (The vertical lines in each
panel mark the radial position where $\alpha=2$.)  Both the MB and LB
density profiles are also different from those found in \citet{hw10},
as those models can have central density cusps. 

We conclude that applying entropy maximization to the LB case, coupled
with a very accurate approximation for $\ln n!$ produces (isotropic)
density profiles that are distinct from any other first-principles
function, or fits to the results of N-body simulations. We will return
to this point in Section~\ref{conc2}.




The MB case is intended to represent collisional systems because the
co-habitation of the phase-space elements is allowed.  The density
profiles with $\nu=100$ and $1000$ are given in Figure~\ref{rho_mb}a.  
The $r_{\rm max}$ scaling distance corresponds to the radius where a 
non-moving particle has an energy for which the distribution function 
disappears. The second panel of that figure (\ref{rho_mb}b) shows the 
slope of the logarithmic density profile $\alpha$ along with curves 
corresponding to three other well-known analytical density profiles.  
These MB solutions are basically identical to King models (see
Figure~\ref{kingcomp}).  This result comes as no surprise in the light
of \citet{m96}, who did not use any approximation for $\ln x!$, and
obtained density profiles very similar to King's. In effect,
abandoning the Stirling approximation and treating the low occupation
number limit with the respect it apparently deserves, demonstrates
that the King distribution function is the result of maximizing
entropy in MB statistics, under the conditions of fixed mass and
energy. 

The consequence of using a very good smooth approximation to $\ln x!$,
as done here, instead of the exact expression which gives discrete
step-like values, as was done in \citet{m96}, is that the former
results in spot-on matches to the King profiles, while the latter
display modest differences from the King profile shape, especially for
low $\Psi(0)/\sigma$ values; see Figure 1 \citet{m96}. 

The value of $\nu$ appears to play a role analogous to the King
scaling factor $\Psi(0)/\sigma^2$, where $\Psi(0)$ is a relative
potential energy at the center, and $\sigma$ is an energy scaling
constant \citep{bt87}.  Increasing $\nu$---or
$\Psi(0)/\sigma^2$---increases the concentration of the system.
Recall that the higher the concentration, the closer the King model is
to an isothermal sphere.

We find that the concentrations of the King models that most closely
resemble the Maxwell-Boltzmann density profiles increase as the value
of $\nu$ increases.  Recall that $\nu$ is the number of fine-grained
micro-cells that occupy a coarse-grained macro-cell, so as $\nu$
increases the distribution function becomes ``grainier''.  At the same
time, increasing the concentration of a King model makes it more
closely resemble an isothermal sphere.  Putting these two observations
together implies that increasing the coarseness of the distribution
function (by increasing the number of fine-grained cells contained in
a coarse-grained cell) should result in densities that have more
isothermal aspects.

Finally, we note that both of the families of profiles obtained here
have flat density cores, while cosmologically simulated halos have
density cusps. 

\subsection{Results of Entropy Production Extremization}\label{conc1b}


Some aspect of simulations may delay, or even disallow, the maximizing
of entropy necessary to achieve thermal equilibrium.  In such a
frustrated case, it is possible that a collisionless system finds a
stationary state by extremizing its entropy production.  As a more
concrete example of this situation, imagine a metal bar with one end
exposed to a blowtorch and the other end in an ice bath.  When the bar
reaches a stationary state with a time-independent, but spatially
varying, temperature distribution, entropy production in the bar is an
extremum \citep[][Ch. 5, Sec. 3]{dgm84}.  We have begun to explore the
possibility that the mechanical equilibria of simulated collisionless
systems can be explained as states of extreme entropy production.

To this end, we have derived expressions for entropy production in
collisionless systems using the Lynden-Bell statistics, and in
collisional systems, using Maxwell-Boltzmann statistics. Further, we
have found the form of the relaxation function required to guarantee
an entropy production extreme.  Recall that the relaxation function is
the right-hand side of the Boltzmann equation,
Equation~\ref{boltzeqn}, that describes the coarse-grained
distribution function evolution.

\begin{figure}
\plotone{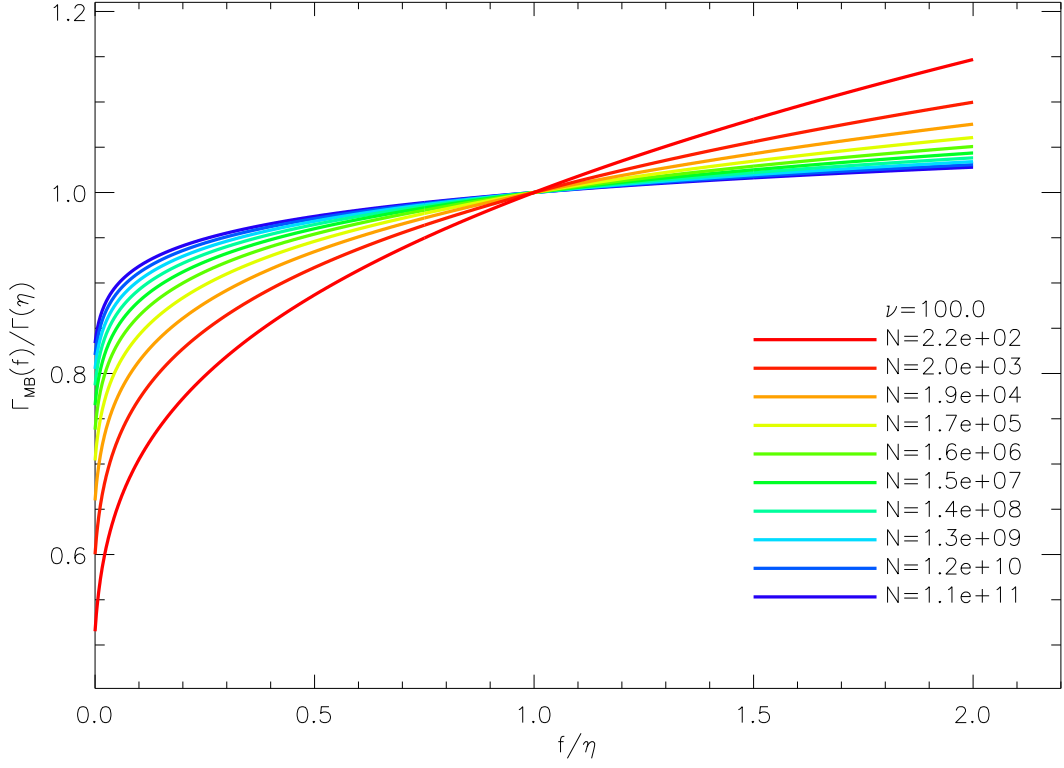}
\figcaption{The behavior of the relaxation function $\Gamma_{\rm MB}$
as a function of the coarse-grained distribution function $f$ in a
system that obeys Maxwell-Boltzmann statistics.  The various lines
represent functions defined with different $N$ values, where $N$ is
the number of phase-space elements in a system.  As with the LB case,
as $N$ increases, the relaxation function becomes more constant.  Note
that the horizontal axis scale differs from that in
Figure~\ref{lbrelax}a since there is no restriction on the value of
the coarse-grained distribution function in the MB case.  All curves
correspond to cases where $\nu=100$.
\label{mbrelax}}
\end{figure}

As in our previous work which was valid only for large phase-space
occupation numbers, the Maxwell-Boltzmann relaxation function
(pictured in Figure~\ref{mbrelax}) is positive for all values of $F$
(as long as $N > 1$) and becomes more constant as the number of
particles $N$ increases.  It is clear that the basic form of the
relaxation function is not dependent on the specifics of the
distribution function.  A visual comparison between
Figure~\ref{mbrelax} and Figure 1 in \citet{bw11} reinforces this
point.

\begin{figure}
\epsscale{0.85}
\plotone{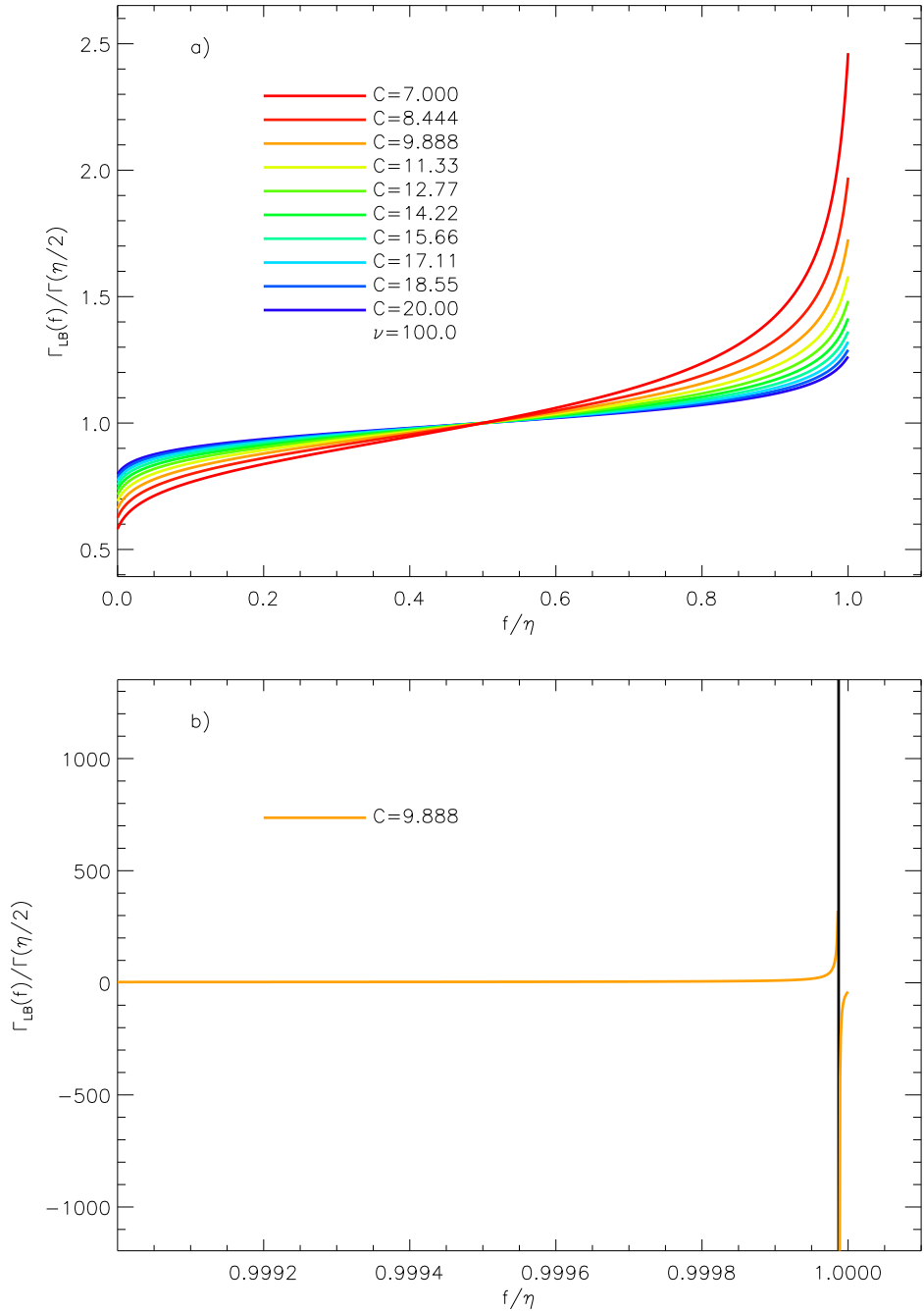}
\figcaption{The behavior of the relaxation function $\Gamma_{\rm
LB}$ as a function of the coarse-grained distribution function $f$ in
a system that obeys Lynden-Bell statistics.  Panel a includes various
lines representing functions defined with different $C$ values, where
$C = \ln{N} - 1 + M/N(\nu \ln{\nu} - \ln{2\pi})$.  $N$ is the number
of phase-space elements in a system, and increasing $N$ results in
increasing $C$.  As $N$ increases, the relaxation function becomes
more constant.  All curves correspond to cases where $\nu=100$.  As
$\nu$ increases, these curves develop singularities near $f=\eta$, as
shown in panel b.
\label{lbrelax}}
\epsscale{0.85}
\end{figure}

The general form of the Lynden-Bell relaxation function shown in
Figure~\ref{lbrelax} has similarities and differences with its large
occupation number version derived in \citet{bw11}.  The similarities
are that both the functions increase with $f/\eta$ up to just short of
$f/\eta=1$, then spike towards large positive values of $\Gamma_{LB}$,
go through a singularity at $f/\eta\simlt 1$, then
become negative, and eventually asymptote to $\Gamma_{LB}=0$ as
$f/\eta\rightarrow 1$ (see the bottom panel of Figure~\ref{lbrelax}).
The difference is that the large occupation number $\Gamma_{LB}$ is
independent of $\nu$, while its arbitrary occupation number analogue
is not. In the latter case the values of $C$ and $\nu$ that produce
the singularity are linked through the relationship,
\begin{equation}
\log{\frac{\nu_{\rm min}}{100}} = 0.4348 (C - 4.1518),
\end{equation}
where $\nu_{\rm min}$ is the minimum value of $\nu$ that will cause a
singularity for a given $C$.  This expression is valid as long as
$\nu_{\rm min} \ga 10$.  The importance of the singularity in  both
cases is that it signals that the relaxation function will reach zero
when the coarse-grained distribution function $f$ reaches the
fine-grained value $\eta$, bringing the evolution of the distribution
function to a halt.  

To sum up, extremizing entropy production in the LB case leads to a
relaxation function $\Gamma_{LB}$ vs. $f/\eta$ shape that drives
$\Gamma_{LB}$, and hence the entropy production rate $\sigma_{Lb}$, to
zero, which in turn means that the endpoint of evolution has a maximum
(more correctly, an extremum) entropy. In other words, the entropy
production extremization procedure that we followed in
Section~\ref{lbsigma} tell us that the final state of a
self-gravitating collisionless system is a state of maximum entropy.
Since in Section~\ref{lbsmax} we derived such a state, using entropy
maximization, it must be the same state. 

The important point in both the LB and the MB case is that extremizing
entropy production leads to relaxation functions that drive a
coarse-grained distribution function to behave like a fine-grained
distribution function.  This has a bearing on the `incompleteness of
relaxation', sometimes alluded to when describing stationary-state
collisionless systems. Our results suggest that if incomplete
relaxation does happen, it is not due to a system reaching an entropy
production extreme.  Rather, it appears that coarse-grained evolution
will proceed until entropy production ceases, when $\Gamma=0$, and so
full relaxation will be achieved in self-gravitating systems. As an
immediate consequence we conclude that entropy maximization is the 
correct---and more direct---procedure to take to arrive at the 
description of steady-state self-gravitating systems.

To test the hypothesis that for self-gravitating systems the state of
entropy production extreme coincides with the state of maximum entropy, 
we are undertaking a comparison between these analytical descriptions of
entropy behavior in collisionless systems and results of $N$-body and
semi-analytical simulations \citep[\eg\ the extended secondary infall
model ESIM][]{w04,a05}.  This further work will help settle exactly
what role entropy production plays in determining collisionless
equilibria.

\subsection{Outstanding Questions}\label{conc2}

Aside from the future work described above, to quantify entropy
production rate in simulated numerical systems, there are also a few
questions that remain unanswered.

\subsubsection{Maximizing Entropy in Phase-space vs. Energy-space}

\cite{hw10} and the present paper maximized entropy, but in different
state-spaces, energy and $(\vx,\vv)$ phase-space, respectively.  The
results of the two studies, for example in terms of the density
profiles, are very different from each other. It is not surprising
that they are different, but it is not immediately obvious which one
of the two approaches would produce a better description of the
results of collisionless N-body simulations. After looking at all the
density profiles it is seen that \citet{hw10} profiles are similar to
those of simulated systems \citep{wh10,whw10}, while those from the LB
case in the present work (Section~\ref{lbsmax}, Figures~\ref{rho_lb}
and~\ref{lbcomp}) are not. 
 
We note that our LB entropy maximization results are what \citet{lb67}
would have obtained had he not used the Stirling approximation.  The
procedure does not rely on any approximations, and is well motivated
from first principles, assuming that all states are equally accessible
and efficient mixing in phase-space can be achieved. This begs the
question, do such systems exist?  More work with simulations is needed
to address this question. 

\subsubsection{The Necessity of the Low Occupation Number Regime}

Our MB entropy maximization results of Section~\ref{smax} \citep[and
the very similar results of][]{m96} describe some globular clusters well
\citep{ehi87,mh97,wbh11}, and the results of \citet{hw10} describe
N-body and ESIM halos.  These facts lead us to the following chain of
reason.  First, globular clusters, along with N-body and ESIM halos,
are self-gravitating systems.  Second, the entropy maximization
procedures that produce successful models of these systems require
correct treatment of low phase-space occupation numbers.  Therefore,
it appears that the low occupation number regime is an important
distinction between self-gravitating and non-self-gravitating systems.

Why is this the case?  We speculate that self-gravitating systems are
special not only because they are finite in extent due to the
long-range nature of the force, but also because of correlations
between spatial regions and energies of particles.  Specifically, the
center of the potential contains phase-space elements with the most
bound energies, while the outskirts are populated by elements with the
least bound, or even zero energies.  These requirements are peculiar
to self-gravitating systems, and apparently necessitate low occupation
numbers because the numbers of particles in these regions must
approach zero.

However, the above hypothesis does not explain why the discrete
version of $\ln x!$ must be smoothed to represent real systems. This
smoothing, though apparently necessary, is still not well justified.

\subsubsection{Should Maximum Entropy Imply Constant Temperature?}

The steady-state of self-gravitating systems can apparently be
calculated as the maximum entropy state. Though these systems are in
mechanical equilibrium, one could argue that they are not in
conventional thermal equilibrium because their kinetic temperature is
manifestly different across the system. Can we reconcile maximum
entropy defining thermodynamic equilibrium with the resulting
non-constant temperature?  (Note that a realistic self-gravitating
system cannot have a constant kinetic temperature, because there is no
finite-mass solution to the Jeans equation with constant $\sigma$.)
This contradiction is part of the reason why it made sense to
conclude, based on \citet{lb67} work that there is no maximum entropy
state for self-gravitating systems. 

We propose that for gravitationally-bound collisionless systems, one
must carefully separate kinetic temperature $T_k$ from thermodynamic
temperature $T_t$.  To be clear, the kinetic temperature we are
referring to is related to the rms value of the peculiar velocity in a
system, $T_k \propto \langle v_p^2 \rangle$.  The thermodynamic
temperature is linked to the energy scale $\beta$ that serves as a
Lagrange multiplier through $\beta = 1/(k_B T_t)$.  By definition,
$T_t$ must be a constant, and there is no contradiction as in
\citet{lb67}---systems in the maximum entropy state are characterized
by a constant thermodynamic temperature, even as they have a varying
kinetic temperature.

We can also address a related question involving the temperature and
energy change that appear in the thermodynamic definition of entropy,
$\dif S = \delta Q/T$.  This temperature must be a constant, so we
would associate this with the thermodynamic temperature $T_t$.
$\delta Q$ (normally called heat) refers to all non-work exchanges of
energy, and so must be replaced by $\delta \epsilon$ because in
collisionless self-gravitating systems what is being transferred is
total energy, potential and kinetic, not just kinetic.  Entropy changes   
can occur due to changes in mass distribution as well as kinetic energy   
transport.

\acknowledgments The authors gratefully acknowledge support from NASA
Astrophysics Theory Program grant NNX07AG86G.  We also thank our
anonymous referee for several helpful suggestions.

\appendix

\section{Maxwell-Boltzmann Statistics}

\subsection{Entropy Maximization}\label{smax}


Using the results of \S~\ref{stat}, the multiplicity function for a
system obeying MB statistics combined with the
definition of entropy (Equation~\ref{s0}) produces,
\begin{equation}\label{smb0}
S_{\rm MB} = \kb \left[ \ln{N!} - \sum_i \ln{n_i!} + N\ln{\nu}
\right],
\end{equation}
where the summation involving the macro-cell occupation $n_i$ runs
over the number of macro-cells.  We will assume that $N \gg 1$, so
that $\ln{N!} = N\ln{N} -N$.  However, we will not use the Stirling
approximation for the second term on the right-hand side of
Equation~\ref{smb0}. With our approximation, Equation~\ref{approxn},
we can now rewrite the entropy as,
\begin{equation}
S_{\rm MB} = S_{\rm MB,0} - \kb \sum_i \left[ (n_i+1/2)\ln{(n_i+1)} +
\lambda_{0,n_i} \right],
\end{equation}
where $S_{\rm MB,0}=N\kb [\ln{(N\nu)}-M\ln{2\pi}/2N]$ is a constant,
and $M$ is the total number of macro-cells.

We now replace the discrete macro-cell occupation number $n_i$ with
the coarse-grained distribution function $f$ to produce
\begin{equation}\label{mbs0}
S_{\rm MB} = S_{\rm MB,0} - \frac{\kb}{\nu \varpi} \iint \left[
\left(\frac{\nu f}{\eta}+1/2 \right) \ln{\left(\frac{\nu f}{\eta}+1
\right)} + \lambda_{0,\nu f/\eta} \right] \: \dif\vx \: \dif\vv.
\end{equation}
To maximize this entropy function, we set $\delta S_{\rm MB}=0$,
subject to constant mass and energy constraints.  The expression that
results is,
\begin{equation}\label{smbmax}
\ln{\left(\frac{\nu f}{\eta} +1 \right)} +\frac{(\nu f/\eta +
1/2)}{(\nu f/\eta + 1)} + \frac{\eta}{\nu} \frac{\partial
\lambda_{0,\nu f/\eta}}{\partial f} + \mu + \beta \epsilon = 0,
\end{equation}
where $\mu$ and $\beta$ are undetermined multipliers associated with
mass and energy conservation, respectively.  The $\epsilon$ term is
the specific energy of a phase element located at position $\vx$ with
velocity $\vv$, $\epsilon=v^2/2 + \Phi$.  Note that if the standard
Stirling approximation had been employed, the second and third terms
would be absent and the logarithmic term would simply read $\ln{(\nu
f/\eta)}$.  In this case, the usual, physically inconsistent, MB
distribution function would result.

\begin{figure}
\plotone{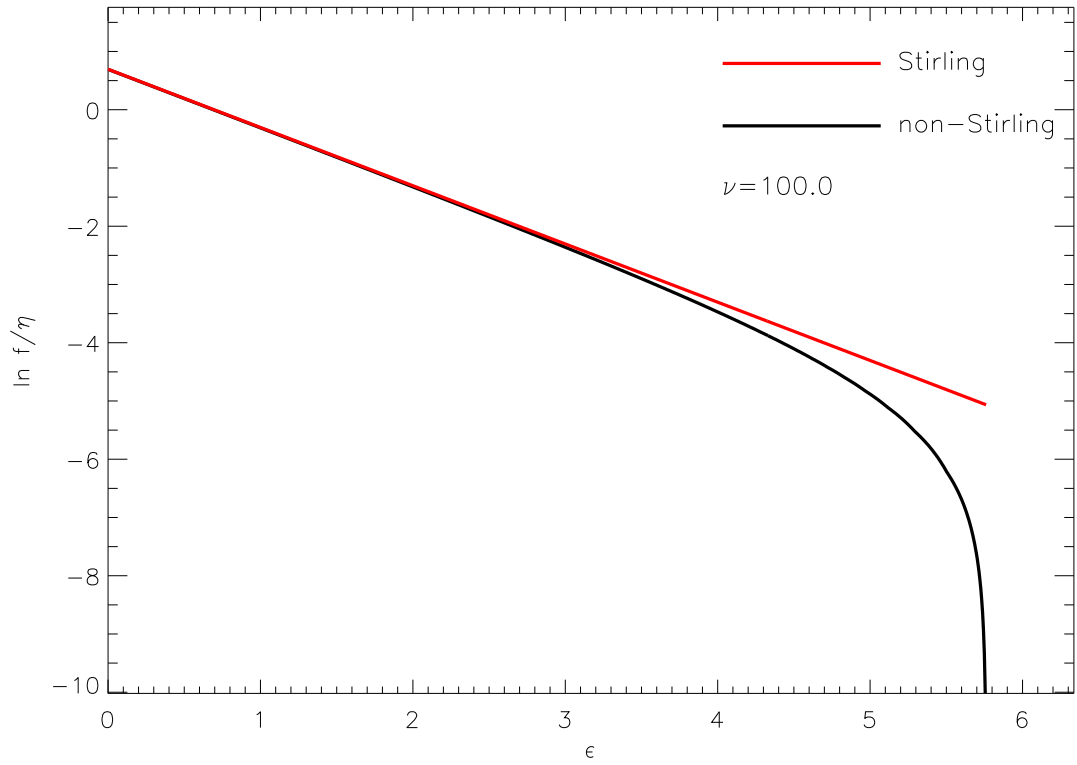}
\figcaption{The natural logarithm of the Maxwell-Boltzmann
distribution function resulting from the non-Stirling approximation.
The Maxwell-Boltzmann function derived using the Stirling
approximation is also shown.  The non-Stirling approximation
distribution function reaches zero for a finite $\epsilon$, unlike the
exponential function derived using the Stirling approximation.
\label{f_mb}}
\end{figure}

\begin{figure}
\plotone{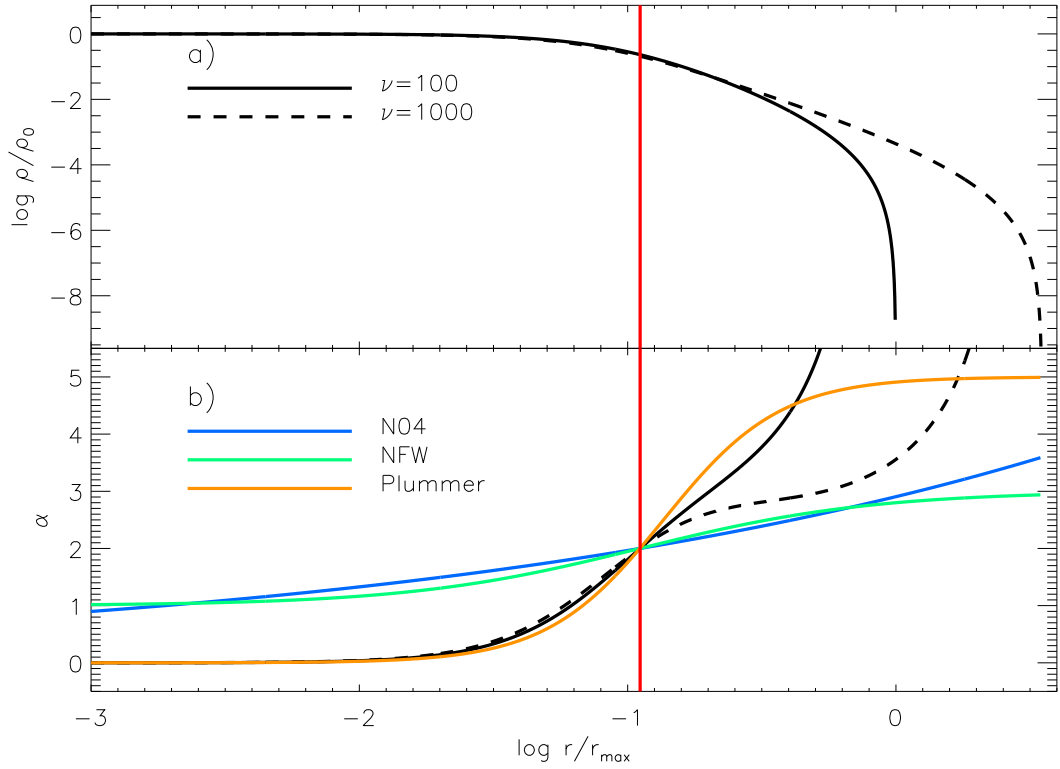}
\figcaption{(a) Logarithmic density and (b) the corresponding
$\alpha=-\dif \log{\rho}/\dif \log{r}$ versus logarithmic scaled
radius for the Maxwell-Boltzmann case.  The thick solid black lines
represent models with $\nu=100$ while the thick dashed black lines
correspond to models with $\nu=1000$.  The vertical lines mark where
the logarithmic slope of the $\nu=100$ model is isothermal
($\alpha=2$).  The $\alpha$ profiles for NFW, \cite{n04}, and Plummer
models are also shown.  As in Figure~\ref{rho_lb}, the $\nu=1000$
curve has been shifted horizontally to align the locations where
$\alpha=2$.  The central cusp of the NFW and \cite{n04} models
($\alpha \rightarrow 1$ as $r \rightarrow 0$) differs markedly from
the core present in the Maxwell-Boltzmann case ($\alpha \rightarrow 0$
as $r \rightarrow 0$).  The presence of the core in the
Maxwell-Boltzmann case is independent of $\nu$.
\label{rho_mb}}
\end{figure}

\begin{figure}
\plotone{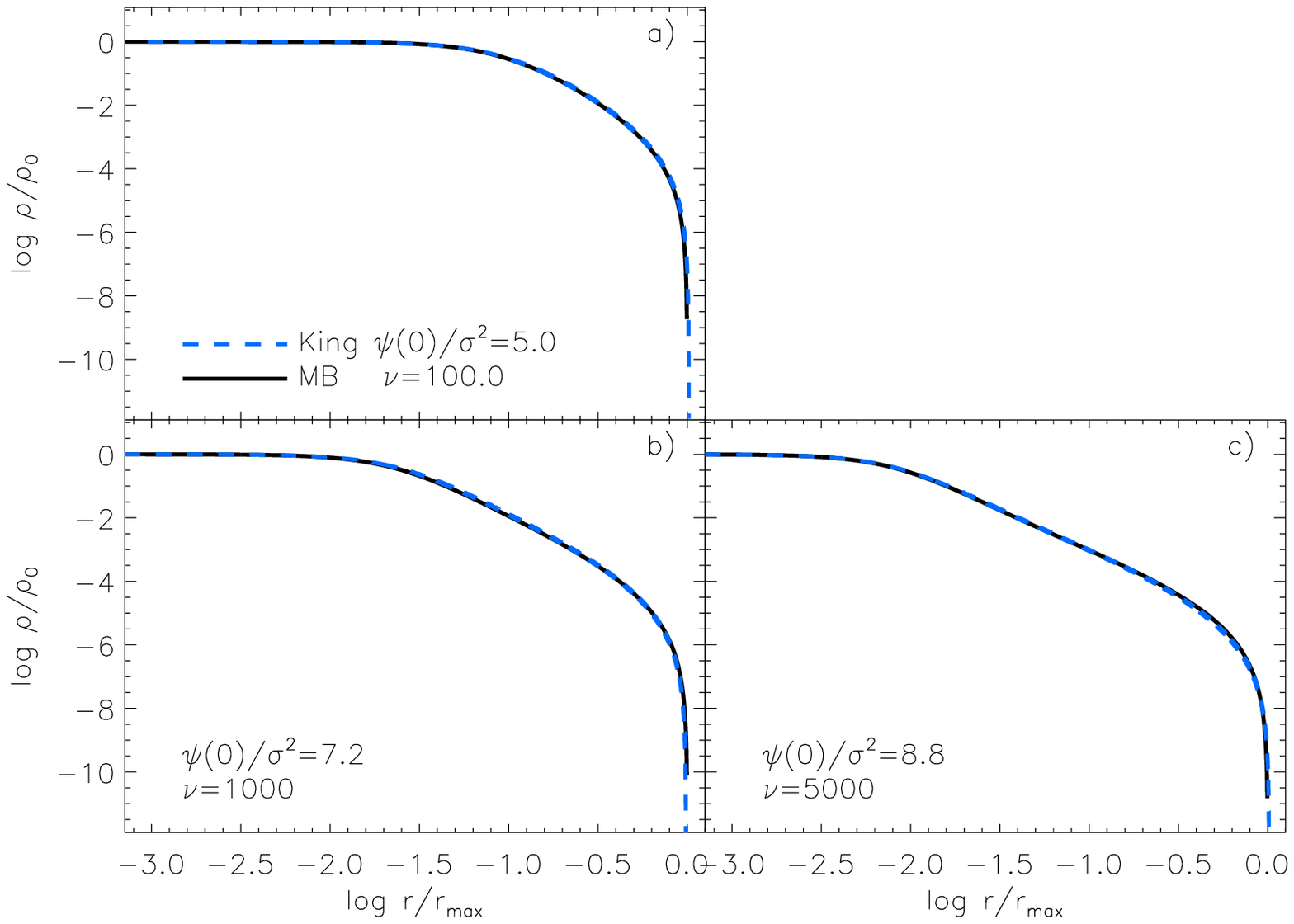}
\figcaption{Comparisons between Maxwell-Boltzmann (solid) and King
(dashed) density profiles for different $\nu$ values; (a) $\nu=100$,
(b) $\nu=1000$, (c) $\nu=5000$.  The King profile scale factor
$\Psi(0)/\sigma^2$ is given in each panel.  The dashed lines represent
King density profiles while the solid lines show the Maxwell-Boltzmann
distributions.  The comparison between these two profile types reveals
more similarity than when the \citet{m96} and King models are
compared.
\label{kingcomp}}
\end{figure}

As before, we make a change of variables, $F \equiv \nu f/\eta$,
and the derivative of the $\lambda$ function is given by 
Equations~\ref{lamderiv}. 
Equation~\ref{smbmax} can then be re-cast as,
\begin{equation}
\ln{(F+1)} - \frac{(264F+276)}{576(F+1)(F+600/576)} + \mu + \beta
\epsilon = 0.
\end{equation}
Again, we have not searched for an analytical solution, but graphical
solutions for $F$ for a range of $\epsilon$ values can be combined to
produce a plot of $F(\epsilon)$.  Specifically, Figure~\ref{f_mb}
illustrates the behavior of the normalized coarse-grained distribution
function $f/\eta$ for a particular value of $\nu$. Values of the 
Lagrange multipliers $\mu$ and $\beta$ were adjusted so that $\epsilon$
is always positive.  Because MB statistics have no exclusion principle, 
$f/\eta$ can have values above 1.

\subsection{Entropy Production Extremization}\label{sigmax}

We begin by writing entropy in terms of entropy density,
\begin{equation}
S=\int \rho s \: \dif\vx,
\end{equation}
where $\rho$ is mass density, $s$ is the specific entropy, and the
integral is taken over the spatial extent of the system.  From the
entropy form given in Equation~\ref{mbs0}, the entropy density can now
be written as,
\begin{equation}\label{mbsdens}
\rho s_{\rm MB} = - \frac{\kb}{\nu \varpi} \int \left[
(F+1/2) \ln{(F+1)} + \lambda_{0,F} -F \frac{S_{\rm MB,0}}{N\kb} 
\right] \: \dif\vv,
\end{equation}
where, as earlier, $F=\nu f/\eta$.

Taking a partial time derivative of Equation~\ref{mbsdens} results in,
\begin{equation}\label{mbk1}
\frac{\partial}{\partial t}(\rho s_{\rm MB}) = -\frac{\kb}{\nu
\varpi} \int \frac{\partial F}{\partial t} \left[
\ln{\left(F+1\right)} + \frac{(F+1/2)}{(F+1)} 
+ \frac{\partial \lambda_{0,F}}{\partial F} -
B \right] \: \dif\vv,
\end{equation}
where $B=S_{\rm MB,0}/N\kb$ is a constant.

Substituting $\partial F/\partial t$ from the Boltzmann equation into
Equation~\ref{mbk1} results in a lengthy expression, similar to 
Equation~\ref{lbk2} in the Lynden-Bell case. We will deal with this
expression term by term.  For reference, the expression after substitution
is,
\begin{eqnarray}\label{mbk2}
\frac{\partial}{\partial t}(\rho s_{\rm MB}) & = & 
-\frac{\kb}{\nu \varpi} \int (-\vv \bcdot \vnab F) \left[ \ln{(F +1)}
+ \frac{(F+1/2)}{(F+1)} + \frac{\partial \lambda_{0,F}}{\partial F} -
B \right] \: \dif\vv -
\nonumber \\
 & & \frac{\kb}{\nu \varpi} \int (-\va \bcdot \vnab_v F) \left[ \ln{(F
+1)} + \frac{(F+1/2)}{(F+1)} + \frac{\partial \lambda_{0,F}}{\partial
F} - B \right] \: \dif\vv +
\nonumber \\ 
& & \frac{\kb}{\nu \varpi} \int \gamma \left[ \ln{(F +1)} +
\frac{(F+1/2)}{(F+1)} + \frac{\partial \lambda_{0,F}}{\partial F} -
B \right] \: \dif\vv,
\end{eqnarray}
where $\gamma=(\nu \Gamma)/\eta$.

The process to evaluate these integrals is very similar to what has
been discussed in the Lynden-Bell case.  
Using the fact that $\vv \bcdot \vnab F=\vnab \bcdot F \vv$, the first
integral on the right-hand side of Equation~\ref{mbk2} can be
re-written as,
\begin{equation}\label{mbk3}
-\vnab \bcdot \int \left[ (F+1/2)\ln{(F+1)} + \lambda_{0,F} - F
B \right] \vv \: \dif \vv.
\end{equation}
Note that this expression is reminiscent of the form of the entropy
density in Equation~\ref{mbsdens}.

We next turn our attention to the terms involving acceleration in
Equation~\ref{mbk2}.  The fact that
\va\ is velocity-independent implies that $\va \bcdot \vnab_v f = \vnab_v
\bcdot \va f$, a fact that will be used often in dealing with these
terms.  Let us start with the first part of the
the second term on the right-hand side of Equation~\ref{mbk2},
\begin{equation}
\int (-\va \bcdot \vnab_v F) \ln{(F+1)} \: \dif\vv = -\int (\vnab_v
\bcdot \va F) \ln{(F+1)} \: \dif \vv.
\end{equation}
Integrating by parts produces two terms, one with the form,
\begin{equation}
\vnab_v \bcdot [\va F \ln{(F+1)}] \: \dif \vv.
\end{equation}
which equals zero after using the divergence theorem and the fact that
any physical distribution function must vanish for large velocities.
The other term resulting from the integration by parts is,
\begin{equation}
\va \bcdot \int \left( \vnab_v F \right) \frac{F}{F+1} \: \dif \vv =
\va \bcdot \int \vnab_v \left[ (F+1) - \ln{(F+1)} \right] \: \dif \vv.
\end{equation}
A single component of this integral will appear as,
\begin{equation}
\int_{v_{1,min}}^{v_{1,max}} \frac{\partial}{\partial v_1} \left[ 
(F+1) - \ln{(F+1)} \right] \: \dif v_1 = (F+1) - \ln{(F+1)}
|_{v_{1,min}}^{v_{1,max}},
\end{equation}
where $v_{1,min} = -v_{1,max}$, representing the maximum speed
possible for the system.  Assuming that $F$ is even in $v_{1,max}$ and
that $F(v_{1,max})=0$ so that the distribution function disappears
at the maximum speed, this integration results in zero.  Similar
manipulations can be applied to the rest of the acceleration-dependent
parts of the second term on the right-hand side of
Equation~\ref{mbk2}, resulting in the entire second term being equal
to zero.  Equation~\ref{mbk2} now has the form,
\begin{eqnarray}\label{mbk4}
\frac{\partial}{\partial t}(\rho s_{\rm MB})& = &
-\frac{\kb}{\nu \varpi} \left\{ -\vnab \bcdot \int \left[ (F+1/2)
\ln{(F+1)} + \lambda_{0,F} - FB \right] \vv \:
\dif\vv + \right. \nonumber \\
& & \left. \int \gamma(F) \left[ \ln{(F+1)} + \frac{F+1/2}{F+1} +
\frac{\partial \lambda_{0,F}}{\partial F} - B
\right] \right\} \: \dif\vv.
\end{eqnarray}
 
Assuming the velocity field to be composed of mean and
peculiar components $\vv = \vvz + \vvp$, we can re-cast
Equation~\ref{mbk4} as,
\begin{eqnarray}\label{mbk5}
\frac{\partial}{\partial t}(\rho s_{\rm MB}) & = &
-\vnab \bcdot \rho s_{\rm MB} \vvz -
\frac{\kb}{\nu \varpi} \Bigg\{
-\vnab \bcdot \int \left[ (F+1/2) \ln{(F+1)} + \lambda_{0,F} -
FB \right] \vvp \: \dif\vv + 
\nonumber \\
 & & \int \gamma(F) \left[ \ln{(F+1)} + \frac{F+1/2}{F+1} +
\frac{\partial \lambda_{0,F}}{\partial F} - B
\right] \: \dif\vv \Bigg\}.
\end{eqnarray}
We now equate the terms in Equation~\ref{mbk5} to those in
the continuous version of the entropy density evolution equation, 
Equation~\ref{s6},
\begin{equation}
\frac{\partial}{\partial t}(\rho s) = -\vnab \bcdot
(\boldsymbol{\Sigma} + \rho s \vvz) + \sigma.
\end{equation}
The entropy flux $\boldsymbol{\Sigma}$ is given by the integral in the
second term on the right-hand side of Equation~\ref{mbk5} and
represents randomly fluxed entropy.  The remaining term is the entropy
production for the system,
\begin{equation}\label{mbsprod}
\sigma_{\rm MB}=
-\frac{\kb}{\nu \varpi} \int \gamma(F) \left[ \ln{(F+1)} + 
\frac{F+1/2}{F+1} + \frac{\partial \lambda_{0,F}}{\partial F} - 
B \right] \: \dif\vv.
\end{equation}
This equation explicitly demonstrates how the non-collisionless nature
of the coarse-grained distribution function leads to changes in
entropy.

As mentioned in \S~\ref{intro}, thermodynamic non-equilibrium systems
can have steady-states described by extrema of entropy production.
We then set $\delta \sigma_{\rm MB} = 0$.  Taking the variation of 
Equation~\ref{mbsprod}, we find
\begin{equation}
\delta \sigma_{\rm MB} = -\frac{\kb}{\nu \varpi} \int \delta F
\left\{ \frac{\dif \gamma}{\dif F} \left[ \ln{(F+1)} +
\frac{F+1/2}{F+1} + \frac{\dif \lambda_{0,F}}{\dif F} - B
\right] + 
\gamma \left[ \frac{2}{F+1} - \frac{F+1/2}{(F+1)^2} +
\frac{\dif^2 \lambda_{0,F}}{\dif F^2} \right] \right\} \: \dif\vv.
\end{equation}
Since $\delta F$ is arbitrary, the variation disappears only when the
term in curly braces is zero.  The condition for an extremum in
entropy production is,
\begin{equation}
\frac{\dif \ln{\gamma}}{\dif F}\left[ \ln{(F+1)} + \frac{F+1/2}{F+1} +
\frac{\dif \lambda_{0,F}}{\dif F} - B \right] + 
\left[ \frac{2}{F+1} - \frac{F+1/2}{(F+1)^2} + 
\frac{\dif^2 \lambda_{0,F}}{\dif F^2} \right]=0.
\end{equation}
The solution for the Maxwell-Boltzmann relaxation function is,
\begin{equation}\label{mbexts}
\gamma_{\rm MB}(F)=P/\left[
\ln{(F+1)} +\frac{F+1/2}{F+1} + \frac{\dif \lambda_{0,F}}{\dif F}
-\frac{S_{\rm MB,0}}{N\kb} \right],
\end{equation}
where the constant can be expressed as 
\begin{equation}
P=\gamma_{\rm MB}(F=\nu) \left[
\ln{(\nu+1)} + \frac{\nu+1/2}{\nu+1} - \frac{\nu+1}{\nu+600/576} - 
(S_{\rm MB,0}/N\kb) \right].
\end{equation}
The $\gamma_{\rm MB}(F=\nu)$ term is the relaxation function value
when the coarse-grained distribution function is equal to the constant
fine-grained distribution function value ($f=\eta$).

\begin{thebibliography}{99}

\bibitem[Aly(1994)]{a94}
Aly, J.-J. 1994, in Lecture Notes in Physics, Vol. 430, Ergodic
Concepts in Stellar Dynamics, ed. V. G. Gurzadyan and D. Pfenniger,
(New York, NY: Springer-Verlag), 226

\bibitem[Austin \etal(2005)]{a05}
Austin, C. G., Williams, L. L. R., Barnes, E. I., Babul, A.,
Dalcanton, J. J. 2005, \apj, 634, 756

\bibitem[Barnes \& Williams(2011)]{bw11}
Barnes, E. I., Williams, L. L. R. 2011, \apj, 728, 136 (BWI)

\bibitem[Binney \& Tremaine(1987)]{bt87}
Binney, J., Tremaine, S. 1987, Galactic Dynamics, (Princeton,
NJ:Princeton)

\bibitem[Chavanis(1998)]{c98}
Chavanis, P. H. 1998, \mnras, 300, 981

\bibitem[de Groot \& Mazur(1984)]{dgm84}
de Groot, S. R., Mazur, P. 1984, Non-Equilibrium Thermodynamics,
(Mineola, NY:Dover)

\bibitem[Elson \etal(1987)]{ehi87}
Elson, R., Hut, P., Inagaki, S. 1987, \araa, 25, 565

\bibitem[Grandy(2008)]{g08}
Grandy, W. T. 2008, Entropy and the Time Evolution of Macroscopic
Systems, (New York, NY:Oxford)

\bibitem[Hjorth \& Williams(2010)]{hw10}
Hjorth, J., Williams, L. L. R. 2010, \apj, 722, 851 (HW10)

\bibitem[Huss \etal(1999)]{h99}
Huss, A., Jain, B., Steinmetz, M. 1999, \apj, 517, 64

\bibitem[Jaynes(1980)]{j80}
Jaynes, E. T. 1980, Annual Review of Physical Chemistry, ed. S.
Rabinovich, (Palo Alto, CA: Annual Reviews)

\bibitem[King(1966)]{k66}
King, I.R. 1966, AJ, 71, 64 (K66)

\bibitem[Lynden-Bell(1967)]{lb67}
Lynden-Bell, D. 1967, \mnras, 136, 101 (LB67)

\bibitem[Madsen(1996)]{m96}
Madsen, J. 1996, \mnras, 280, 1089 (M96)

\bibitem[Meylan \& Heggie(1997)]{mh97}
Meylan, G., Heggie, D.C. 1997, \araa, 8,1

\bibitem[Navarro \etal(1996)]{nfw96} 
Navarro, J. F., Frenk, C. S., White, S. D. M. 1996, \apj, 462, 563

\bibitem[Navarro \etal(2004)]{n04}
Navarro, J. F., Hayashi, E., Power, C., Jenkins, A. R., Frenk, C.
S., White, S. D. M., Springel, V., Stadel, J., Quinn, T. R. 2004,
\mnras, 349, 1039

\bibitem[Ogorodnikov(1957)]{o57}
Ogorodnikov, K.F. 1957, Soviet Astronomy, 1, 748

\bibitem[Plummer(1911)]{p11}
Plummer, H. C. 1911, \mnras, 71, 460

\bibitem[Prigogine(1961)]{p61}
Prigogine, I. 1961, Thermodynamics of Irreversible Processes, (New
York, NY: Interscience)

\bibitem[Schuster(1883)]{s83}
Schuster, A. 1883, British Assoc. Report, 427

\bibitem[Simons(1994)]{s94}
Simons, S. 1994, Am. J. of Phys., 62, 515

\bibitem[Stiavelli \& Bertin(1987)]{sb87}
Stiavelli, M., Bertin, G. 1987, \mnras, 229, 61

\bibitem[Tremaine, H\'{e}non, \& Lynden-Bell(1986)]{t86}
Tremaine, S., H\'{e}non, M., Lynden-Bell, D. 1986, \mnras, 219, 285

\bibitem[van Albada(1982)]{va82}
van Albada, T. S. 1982, \mnras, 201, 939

\bibitem[White \& Narayan(1987)]{wn87}
White, S. D. M., Narayan, R. 1987, \mnras, 229, 103

\bibitem[Williams, Babul, \& Dalcanton(2004)]{w04}
Williams, L. L. R., Babul, A., Dalcanton, J. J. 2004, \apj, 604, 18

\bibitem[Williams \& Hjorth(2010)]{wh10}
Williams, L.L.R. \& Hjorth, J. 2010, \apj, 722, 856

\bibitem[Williams \etal(2010)]{whw10}
Williams, L.L.R., Hjorth, J. \& Wojtak, R. 2010, \apj, 725, 282

\bibitem[Williams \etal(2011)]{wbh11}
Williams, L.L.R., Barnes, E.I. \& Hjorth, J. 2011, \mnras, submitted

\end{thebibliography}
\end{document}